\documentstyle[12pt]{article}
\begin{document}

\begin{center}

             {\bf   IMPLICATIONS OF A QUANTUM MECHANICAL TREATMENT 
                        OF THE UNIVERSE   }\\
\mbox{}\\

		      Murat \"Ozer  \\

			Department of Physics, College of Science\\
				  King Saud University\\
                        P.O. Box 2455, Riyadh 11451, Saudi Arabia \\                    
\mbox{}\\
				{\bf  ABSTRACT}
\end{center}

We attempt to treat the very early Universe according to quantum mechanics. Identifying the scale factor of the Universe with the width of the wave packet associated  with it ,we show that there cannot be an initial singularity and that the Universe expands. Invoking the correspondence principle, we obtain the scale factor of the Universe and  demonstrate that the causality problem of the standard model is solved.

\newpage

Whether the scale factor $a(t)$   of the Universe vanishes at  $t=0$ and  its time dependence are of paramount importance in cosmology based on classical general relativity and its extensions. In the hot big - bang model, the so called standard model, the Einstein field equations 
\begin{equation} 
R_{\mu\nu}-\frac{1}{2}g_{\mu\nu}R_-\lambda g_{\mu\nu}=-8\pi GT_{\mu\nu} ,
\end{equation}
predict an initial singularity. In other words, at $t=0$ $a(t)=0$  due to which physical quantities such as temperature, energy density, and pressure become infinitely large. In the standard model the functional dependence of the scale factor on time, which is
\begin{equation} 
a_{SM}(t)=const.t^{1/2}
\end{equation}					    			     
in the very early Universe is responsible for the causality (horizon) problem [1]. It is hoped by the Cosmologists that the answer to the initial singularity problem and a better understanding of the other cosmological problems  lie in the quantum domain. At present, a complete theory of quantum cosmology [2] does not exist because of lack of a complete theory of quantum gravity. In its present form quantum cosmology aims at obtaining the wave function of the Universe [3,4] by solving the Wheeler - De Witt equation [5], which is the cosmological analogue of the Schr\"odinger equation. Such a wave function of the Universe does not, and cannot, depend on the time explicitly because of general covariance. Thus there is no way of extracting the scale factor of the Universe as a function of the time from such a wave function of the Universe.

The purpose of this letter is to attempt to treat the very early Universe according to quantum mechanics. We contemplate that the spacetime is divided into two regions. At $t=0$  the Universe is centered, within quantum fluctuations, at the origin of a three dimensional flat space, which we choose to call the   " outer space ". The outer space should not be confused with the curved  " inner space ".( We recall that dividing the spacetime into two or more regions with different curvature properties has been employed in more or less similar situations .See, for example, ref.[6] for earlier and ref.[7] for recent works.) The forces of nature act on the constituents of the Universe in the inner space. The only entity in the outer space is the Universe itself. Hence the Universe acts as a free particle in the outer space. To describe the motion of this free particle, the Universe, according to quantum mechanics we thus associate a wave packet to it from its creation until the quantum mechanical behavior ceases. \footnote{We do not know  for sure if  quantum theory can be applied to the present Universe. However, the general approach currently  is  to apply quantum theory to the entire Universe at all times.} But the question " What kind of wave packet describes the quantum mechanical motion of the Universe ? "  must be answered. We are guided by the correspondence principle in our search for the answer to this question. It must be a wave packet that mimics the classical evolution very well in the limit $\hbar\rightarrow 0$, where $\hbar=h/2\pi$  and $h$ is the Planck constant. Therefore the best choice for the wave packet of the Universe is the Gaussian wave packet.(See ref.[8] for a demonstration of the fact that the Gaussian wave packet describes the classical evolution of a free particle very well in this limit.) The essential point in such a treatment is the identification of the scale factor of the Universe with the width of the wave packet (see later). Let $r=|\vec{r}|$  be the position of the Universe in the outer space and $p_{r}=|\vec{p_{r}}|$  be the corresponding conjugate momentum. The wave packet is initially centered, within quantum fluctuations, at $r=0$  and $\langle\vec{r}\rangle=\langle\vec{p_{r}}\rangle=0$. Thus the three - dimensional free minimum wave packet (the Gaussian wave packet) describing the motion of the Universe at $t=0$ is  
\begin{equation} 
\psi(r,0)=\left[\frac{2\pi}{3}(\Delta r)_{0}^{2}\right]^{-3/4}exp\left[-\frac{3r^{2}}{4(\Delta r)_{0}^{2}}\right] .
\end{equation}
At $t=0$, the Universe obeys  the minimum uncertainty relation
\begin{equation} 
(\Delta r)_{0}(\Delta p_{r})_{0}=3\frac{\hbar}{2} ,
\end{equation}			                 			     where $(\Delta r)_{0}$ and $(\Delta p_{r})_{0}$  are the initial uncertainties in the position and the momentum of the Universe. Subsequently, the wave packet in Eq.(3) moves uniformly and at the same time spreads [9] so that the equation of motion of its width is
\begin{equation} 
(\Delta r)_{t}^{2}=(\Delta r)_{0}^{2}+\frac{(\Delta p_{r})_{0}^{2}}{m^{2}}t^{2} ,
\end{equation}				              			     
where $m$ is the mass of the Universe. Next, we  invoke the correspondence principle according to which the wave packet evolves like a classical particle, and hence identify the spread of the wave packet and its width with the expansion of the Universe and its scale factor. Thus  we conclude from Eq.(5) that the scale factor $a(t)$  is given by  
\begin{equation} 
a(t)^{2}=a_{0}^{2}+\gamma t^{2} ,
\end{equation}				  				      where $a_{0}=(\Delta r)_{0}$ and $\gamma=(\Delta p_{r})_{0}^{2}/m^{2}$   \footnote{The scale factor in Eq.(6) with was obtained from a different  motivation in ref.[10]. The same equation with a general  was considered in ref.[11]}.
 We thus  learn two important facts about the Universe from such a quantum mechanical treatment: (1) The Universe must have been created with a very small but nonzero spatial size  to comply with the uncertainty principle. Hence there is no initial singularity. (2) The Universe expands.

A definitive test of  this treatment of the Universe and the identification of the width of the wave packet with the scale factor of the Universe is the success with which deductions from it solve some of the cosmological problems of the standard model. Firstly, there is no initial singularity. Secondly, the Universe expands with the same velocity in all directions. Hence when applied to the inner space this leads to the conclusion that the Universe  is isotropic and thereby homogeneous. Thirdly, the expansion law in Eq.(6) solves the causality (horizon) problem of the standard model in the curved inner space. This inner space, where the forces of nature act, being  homogeneous and isotropic, is thus described by the Robertson - Walker metric, 
\begin{equation} 
ds^{2}=-c^2dt^{2}+a(t)^{2}\left[\frac{dr^{2}}{1-kr^{2}}+r^{2}(d\theta^{2}+sin^{2}\theta d\phi^{2})\right] ,
\end{equation}		           
where $k$ is the curvature constant. The Einstein field equations (1) combined with Eq.(7) yield the Friedman equation 
\begin{equation} 
H^{2}=\left(\frac{\dot{a}}{a}\right)^{2}=\frac{8\pi G}{3}\rho_{everything}-\frac{k}{a^{2}} ,
\end{equation}					
where $\rho_{everything}=\rho_{radiation}+\rho_{\lambda}+\cdots$  is the sum of the energy densities of  everything that acts inside the Universe at and after $t=0$. A very plausible assumption is that $\rho_{everything}$  is positive definite. Then Eq.(8) at $t=0$  with $H=0$  (which follows from Eq.(6) ) yields $k=+1$  so that its right hand side vanishes. Hence we have a solid prediction for the inner three - geometry of the Universe.

 The horizon size $d_{H}(t)$ at time $t$   is the proper distance  traveled by  light emitted at $t=0$:
\begin{equation} 
d_{H}(t)=ca(t)\int_{0}^{t}\frac{dt'}{a(t'}=\frac{c}{\sqrt{\gamma}}a(t)sinh^{-1}\left[\left(\frac{\gamma}{a_{0}^{2}}\right)^{1/2}t\right] , a_{0}\neq 0 ,
\end{equation}                        	      
where $c$ is the speed of light. For the Universe to be causally connected globally at time $t$ it is necessary that $d_{H}(t)\geq d_{P}(t)$  where $d_{P}(t)$  is the proper distance between the galaxy at $r=0$ and the galaxy most distant from it, that is the one at $r_{max}=1$:
\begin{equation} 
d_{P}(t)=a(t)\int_{0}^{r_{max}=1}\frac{dr}{\sqrt{1-kr^{2}}}=a(t)\frac{\pi}{2}  ,
\end{equation} 				   
where $k=+1$ has been used. Thus global causality  is established at   $t=\frac{a_{0}}{\sqrt{\gamma}}$ [10,11]. Therefore the Universe consists of one causally connected region. In Figure 1 we depict the behavior of $d_{H}(t)$ and $d_{P}(t)$  against the cosmic time $t$. This is to be compared with the situation in the standard model  in which the horizon size is given, in the very early radiation - dominated Universe, by
\begin{equation} 
d_{H}(t)=2ct .
\end{equation} 							             Using the fact that entropy is conserved in the standard model, the constant in Eq.(2) is found, in units with, $c=\hbar=k_{B}=1$ to be, $const.=(16g_{eff}\pi^{3}G/45)^{1/4}a_{p}T_{p}$, where $g_{eff}\approx 100$ is the effective number of particle degrees of freedom in the early Universe and $a_{p}\approx 10^{26}m$ and $T_{p}=2.7K$  are, respectively, the present value of the size of the visible Universe and the temperature of the background radiation. The causality is established when $t=(const./2c)^{2}\approx 2.7x10^{7}yr$ which extends well into the matter - dominated era. The Universe in the standard model was then still causally disconnected at the end of the radiation - dominated era! As is well known, this is due to the form of the scale factor in Eq.(2). 

We refer the reader to refs.[10] and [11] for a discussion of a Universe whose scale factor is given by that in Eq.(6) in the very early Universe.

In conclusion, we have attempted to treat the Universe according to quantum mechanics. Identifying the scale factor of the Universe with the width of the wave packet associated with it leads to the conclusion that there is no initial singularity and  that the Universe inevitably expands. When used in the Friedman equation, the scale factor obtained from quantum mechanical considerations and the positive definiteness of the energy density yield $k=+1$ for the three - geometry. Most important of them all,  the causality (horizon) problem of the standard model does not exist with this scale factor.

{\bf Acknowledgements}

We thank Prof. Mahjoob O. Taha for invaluable discussions and a critical reading of the manuscript.

{\bf References}

1. For a review of the cosmological problems in the standard model see 
R. H. Branderberger, Rev. Mod. Phys. {\bf 57} ,1 (1985).\\ 
2. A bibliography of papers on quantum cosmology is J. J. Halliwell, Int. J. Mod. Phys. {\bf A5}, 2473 (1990). \\
3. J.B.Hartle and S.W.Hawking, Phys.Rev. {\bf D28} , 2960 (1983).\\ 
4. A.Vilenkin, Phys.Rev. {\bf D33}, 3560 (1986), ibid {\bf D39},1116 (1989).\\ 
5. J.A.Wheeler, in Battelle Rencontres, edited by C . De Witt and J.A.Wheeler\\ 
(Benjamin, New York 1968); B. S. De Witt, Phys.Rev. {\bf 160}, 1113 (1967).\\ 
6. S. Coleman and F. De Luccia , Phys.Rev. {\bf D12}, 3305 (1980).\\
7. M. Bucher, A. S. Goldhaber , and N. Turok, Phys.Rev. {\bf D52}, 3314 (1995).\\
8. J. V. Narlikar and T. Padmanabhan, Gravity, Gauge Theories and Quantum 
 Cosmology (D. Reidel Publishing Company, Dordrecht, Holland ,1986) p. 29.\\
9. For  the  motion  of  a  wave  packet   see,   for  example,  the  textbook V.  Merzbacher, Quantum Mechanics (John Wiley \& Sons).\\
10. M. \"Ozer and M.O.Taha , Phys.Lett. {\bf B171},363 (1986) ; Nucl.Phys. {\bf B287} ,776 (1987).\\  
11. A - M . M. Abdel - Rahman, Phys.Rev. {\bf D45} , 3497 (1990).

{\bf Figure Captions:}

Figure 1: Plots of the particle horizon $d_{H}/a_{0}$ for $\sqrt{\gamma}=c=1$ (see Eq.(9)) and the proper distance $d_{P}/a_{0}$ between two galaxies at $r=0$ and $r_{max}=1$ versus $t/a){0}$, where $a_{0}$ is the initial size of the Universe at t = 0 (see Eq.(10)).

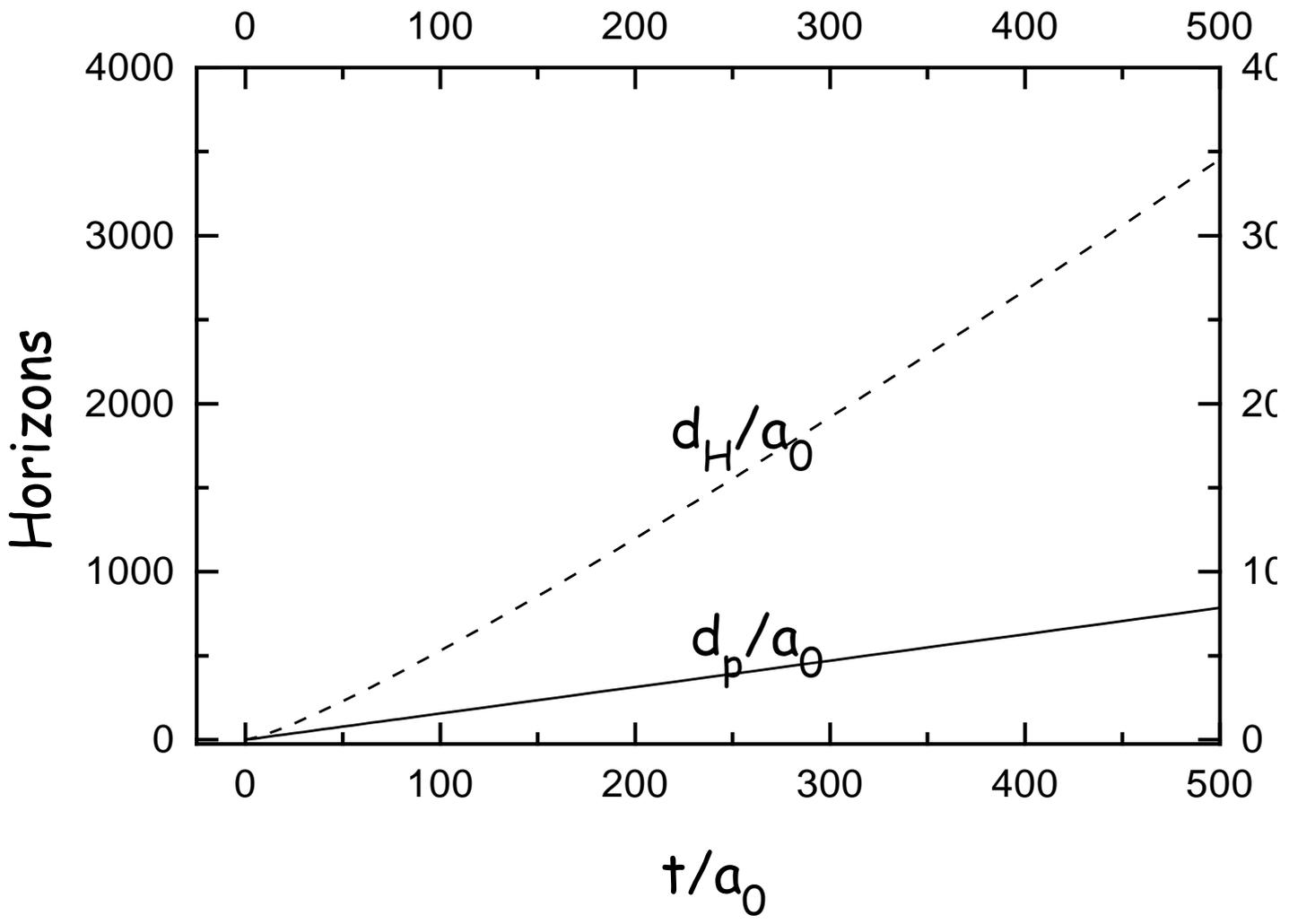
\end{document}